\documentclass[12pt]{article}

\usepackage[utf8]{inputenc} 
\usepackage[T1]{fontenc}    
\usepackage{hyperref}       
\usepackage{url}            
\usepackage{booktabs}       
\usepackage{amsfonts}       
\usepackage{nicefrac}       
\usepackage{microtype}      
\usepackage{lipsum}		
\usepackage{graphicx}
\usepackage[super]{natbib}
\usepackage{doi}

\usepackage{multirow} 
\usepackage{amsmath} 
\usepackage{authblk} 
\usepackage{flafter} 

\usepackage{pdfpages}

\makeatletter
\newcommand*{\citenormal}[2][]{
  \begingroup
  \let\NAT@mbox=\mbox
  \let\@cite\NAT@citenum
  \let\NAT@space\NAT@spacechar
  \let\NAT@super@kern\relax
  \cite[#1]{#2}%
  \endgroup
}
\makeatother

\title{Massively parallel quantum chemistry: \\PFAS on over 1 million cloud vCPUs}

\author[1]{Alan E. Rask}
\author[1]{Lee Huntington}
\author[1]{SungYeon Kim}
\author[1]{David Walker}
\author[1]{Andrew Wildman}
\author[1]{Rodrigo Wang}
\author[1]{Nicole Hazel}
\author[1]{Alan Judi}
\author[1]{James T. Pegg}
\author[1]{Punit K. Jha}
\author[1]{Zara Mayimfor}
\author[2]{Carl Dukatz}
\author[2]{Hassan Naseri}
\author[3]{Ilan Gleiser}
\author[3]{Maxime R. Hugues}
\author[1,4]{Paul M. Zimmerman}
\author[1]{Arman Zaribafiyan}
\author[1,$\dag$]{Rudi Plesch}
\author[1,$\ddag$]{Takeshi Yamazaki}

\affil[1]{Good Chemistry Company, 200-1285 West Pender Street, Vancouver, V6E 4B1, BC, Canada}

\affil[2]{Accenture, 1 Grand Canal Quay, Grand Canal Dock, Dublin, D02 P820, Ireland}

\affil[3]{Amazon Web Services, 410 Terry Ave N, Seattle, 98109, Washington, United States}

\affil[4]{Department of Chemistry, University of Michigan, Ann Arbor, 48109, Michigan, United States}

\affil[$\dag$]{rudi@goodchemistry.com}
\affil[$\ddag$]{takeshi@goodchemistry.com}

\date{}

\begin{document}
\maketitle

\begin{abstract}
	Accurate solutions to the electronic Schr\"odinger equation can provide valuable insight for electron interactions within molecular systems, accelerating the molecular design and discovery processes in many different applications. However, the availability of such accurate solutions are limited to small molecular systems due to both the extremely high computational complexity and the challenge of operating and executing these workloads on high-performance compute clusters. This work presents a massively scalable cloud-based quantum chemistry platform by implementing a highly parallelizable quantum chemistry method that provides a polynomial-scaling approximation to full configuration interaction (FCI). Our platform orchestrates more than one million virtual CPUs on the cloud to analyze the bond-breaking behaviour of carbon-fluoride bonds of per- and polyfluoroalkyl substances (PFAS) with near-exact accuracy within the chosen basis set. This is the first quantum chemistry calculation utilizing more than one million virtual CPUs on the cloud and is the most accurate electronic structure computation of PFAS bond breaking to date.
\end{abstract}

\section*{Introduction}

Quantum chemistry exhibits a great potential to tackle some of humanity's most challenging problems by accurately simulating chemical phenomena, such as drug discovery, efficient solar panel development, finding better battery materials, and removal of toxic substances from the environment. In general, quantum chemical approaches simulate these phenomena by solving the Schr\"odinger equation\cite{Pople_1999,Friesner_2005,Rezac_2016}. However, calculating exact solutions to this problem is only possible for the smallest molecular systems in small basis sets due to extremely high computational demand\cite{Head-Gordon:2008}. Many of the leading chemical problems involve systems much too large for this exact treatment. As an example, polyfluoroalkyl substances (PFAS) are persistent pollutants that have been dubbed ``forever chemicals'' because of their environmental resilience and accumulation in both humans and the environment\cite{evich2022per,crone2019occurrence}. One of the most common PFAS in the environment, perfluorooctanoic acid (PFOA), requires approximately $10^{151}$ determinants to capture the exact energy with full configuration interaction (FCI) when correlating 150 electrons in 330 orbitals using a double-zeta quality basis set, while the largest fully variational calculations to date have considered a maximum of $\sim10^{12}$ determinants\cite{vogiatzis2017pushing}.

To tackle problems at this scale, approximations to the exact solution must be made; whereas, studying fundamental chemical phenomena, such as bond breaking, require highly accurate treatments\cite{NPE_3}. PFAS are especially challenging due to the abundance of fluorine atoms in close proximity, with each successive CF$_2$ adding an additional 18 valence electrons (12 of which are lone pairs on the fluorine atoms). Additionally, fluorine's electronegativity produces strong C$-$F bonds, setting these species apart from systems typically studied with electronic structure methods and requiring a thorough investigation of the dissociation manifold. The complexity of PFAS suggests that conventional computational methods, such as coupled-cluster theory, are not expected to accurately determine physical properties of these species, and therefore more accurate methods are necessary. In order to address these problems, a method that is both scalable and highly accurate must be identified.

The incremental full configuration interaction (iFCI) method is a novel approach to tackle large-scale electronic structure problems with high accuracy\cite{Zimmerman:2017ab, Zimmerman:2017aa, Rask_TM, Rask_porphyrin}. iFCI begins with a many-body decomposition of the energy of a molecular system within a localized orbital basis. This energy decomposition, when truncated, reduces the problem to polynomial scaling, enabling simulation of much larger systems. Furthermore, the method is systematically improvable by choosing a higher order of truncation, and does not require selection of an active space. Even with the efficiency of iFCI, a system such as PFOA would require approximately $10^6$ many-body terms (with an average of about $10^8$ determinants per term in a double-zeta quality basis set). With an estimated average runtime of 15 minutes per term, this calculation would take approximately 28.5 years, solved sequentially. Fortunately, each term at each order of the expansion is completely independent of the others, making iFCI an extremely parallelizable and scalable method. To reduce the time required for the calculation mentioned above to the scale of several hours, approximately $6\times10^5$ iFCI terms must be simultaneously calculated. Roughly 1 million processing units (vCPU) are required to calculate the iFCI energy for the aforementioned problem within the timescale of less than a day. In order to apply iFCI to systems such as PFOA, an efficient implementation of iFCI must be created that can leverage this high degree of parallelism.

When looking towards parallelizing electronic structure methods, most implementations have traditionally targeted supercomputing clusters. Computations require the unique features of these clusters, namely low-latency and high-bandwidth interconnecting networks and shared, parallel file systems. However, supercomputing facilities are costly, and high demand for the existing resources results in long wait times. Furthermore, one million vCPUs are typically not available simultaneously in a single supercomputer. In contrast, commodity cloud computing has seen a surge in popularity over the last few decades due to the low cost of and rapid access to computing power\cite{Durao_2014, Rashid_2018}. Because of the benefits of cloud computing, and because iFCI does not have tightly-coupled parallelism and consequently does not require the unique features of supercomputing clusters, the implementation of iFCI presented in this paper will be fully cloud native, targeting commodity hardware.

In this work, we describe a massively parallel implementation of iFCI that can make use of more than one million vCPUs simultaneously. We discuss the specific technologies that enable commodity cloud computing to run massive computing tasks at this scale. We also address some of the architectural decisions that were driven by iFCI. Lastly, we demonstrate that this implementation can reach that scale while tackling problems of scientific importance.

Due to the global impact and health risks of PFAS and the need for their remediation\cite{evich2022per,crone2019occurrence,ahrens2011polyfluoroalkyl,Accenture_inPrep}, analyzing the bond breaking process of PFAS molecules was selected for these iFCI calculations. Specifically, three representative PFAS molecules -- trifluoroacetic acid, (TFA) perfluorobutanoic acid, (PFBA) and PFOA -- were chosen as target systems based on their environmental relevance and the difficulty of studying these systems with traditional electronic structure theory\cite{Accenture_inPrep}. In particular, capturing the energy required to break the C$-$F bond is important for understanding natural and induced PFAS degradation pathways, as well as designing potential catalysts for PFAS remediation. In the present study, we investigate the rigid-body stretch of a C$-$F bond, which is calculated at a high level of accuracy for each molecule while also demonstrating the scalability of this platform. This is the very first quantum chemistry calculation to utilize more than one million vCPUs on the cloud, and it is the most accurate electronic structure computation of PFAS bond dissociation within the Born-Oppenheimer approximation to date.

\section*{Results}
\subsection*{QEMIST Cloud architecture design}

The massively parallel framework built for the iFCI method was implemented within QEMIST Cloud (https://goodchemistry.com/qemist-cloud/), a computational chemistry software as a service (SaaS) that provides accurate methods to predict the electronic structure and properties of chemical systems with both traditional \textit{ab initio} quantum chemistry and machine learning methods. For this work, our software was implemented using Amazon Web Services (AWS) as the platform of cloud-based computing resources, using Amazon Elastic Compute Cloud (Amazon EC2) Instances with Intel Xeon Scalable processors. 

Before describing the technologies used to implement this software architecture at scale, we will first introduce some terminology. An availability zone is a set of one or more data centers in a particular region. Nodes and instances refer to discrete allocations of hardware resources, and workers refer to processes that use some or all of these resources. A cluster is a collection of nodes, and a spot instance is some spare capacity on AWS that is available at up to a 90\% discount compared to On-Demand prices, but may be reclaimed by AWS before the calculation is finished (with a warning issued two minutes prior to reclamation). Individual compute tasks were performed with standardized units of software, referred to as containers. Kubernetes performed distribution and execution of the compute tasks and containers\cite{brewer2015kubernetes}. Kubernetes was deployed using the AWS managed service, Amazon Elastic Kubernetes Serice.

The architecture supporting the iFCI implementation is structured around the independence and natural parallelism of the subproblems generated by the method of increments\cite{Zimmerman:2017ab}. In iFCI, the $n$-body correlation energy is composed of a summation of independent calculations, each correlating unique sets of $n$ occupied orbitals. Due to the natural independence of each calculation and very low communication between calculations, traditional message passing interface parallelism is not required for these calculations. Instead, when our platform receives an iFCI calculation request, a dedicated decomposition worker is assigned and generates all required subproblems for the current $n$. The decomposition worker then submits subproblems to solver queues, where solver workers pick up and complete each subproblem, writing their results into a central database once finished. The number of solver workers scales dynamically so that larger calculations have more subproblems being solved simultaneously. Once all subproblems for a given $n$ are completed, the decomposition worker sums the total $n$-body correlation energy and moves to the next subsequent $n$ until the desired highest order of $n$ is computed (usually not higher than $n=4$).

\begin{figure}[ht]
\centering
\includegraphics[width=0.8\textwidth]{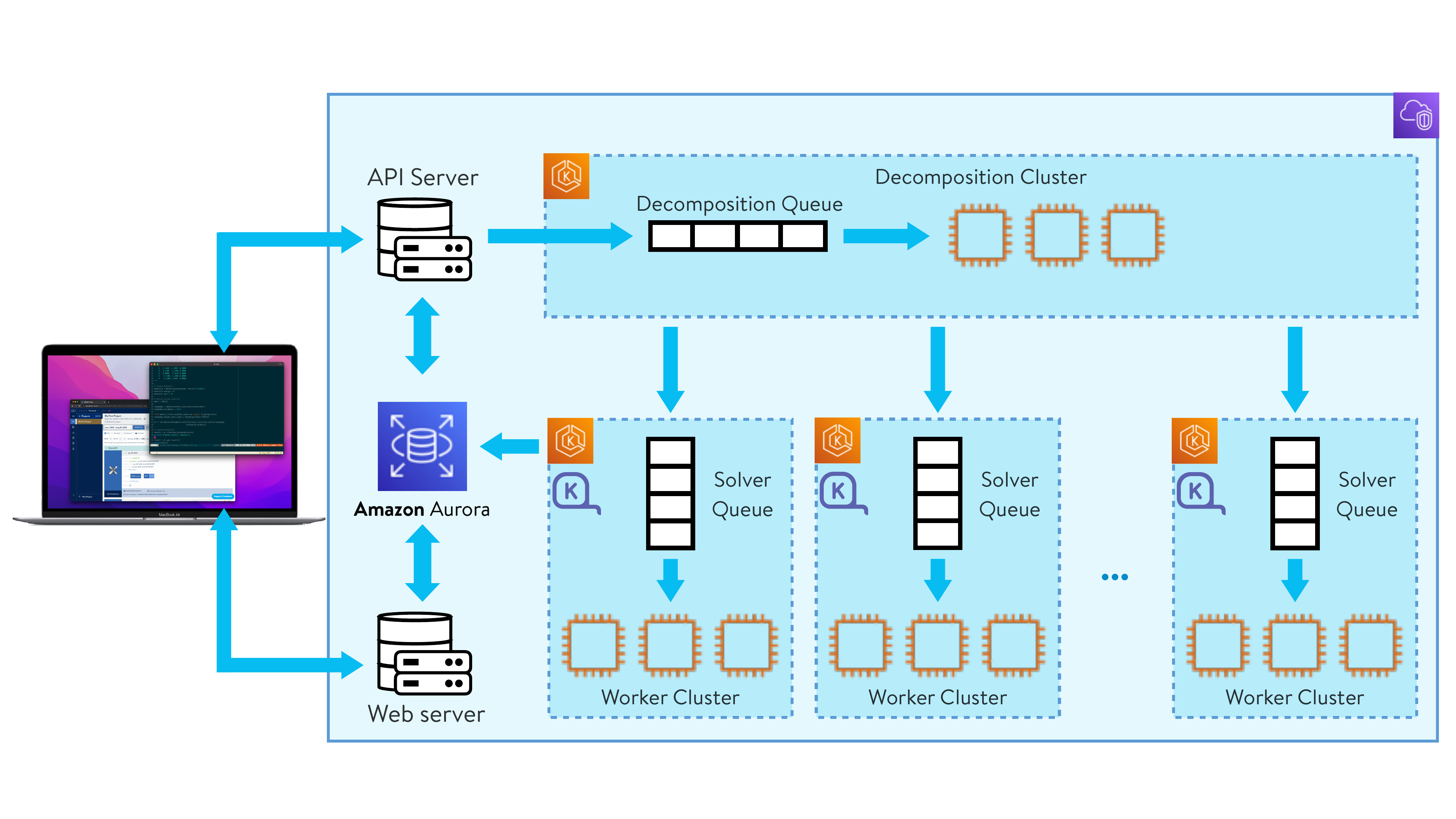}
\caption{QEMIST Cloud architecture overview. The laptop represents a client, stacked white boxes represent queues, and orange CPU symbols represent individual workers. Each dashed box is a group of components that was deployed into a specific cluster.} 
\label{fig:arch}
\end{figure}

To reach the scale required for the largest PFAS in this study (1 million simultaneous vCPUs, or 62,500 workers with 16 vCPUs each), this infrastructure required numerous enhancements. In general, these improvements fell into three categories: more efficient node recruitment, enhanced database connections, and widening of the network to accommodate a large number of nodes. A summary of the final architecture is given in Figure~\ref{fig:arch}.

In order to reach a given scale, enough nodes must be available. While cloud infrastructure providers generally have more than enough to reach the scale aimed for in this work, no one availability zone is guaranteed to have the required amount. To ensure adequate node availability, the solver queues and solver workers were moved from a single cluster into multiple clusters, which distributed the work of provisioning nodes and enabled node recruitment on a larger scale. All clusters were allowed to provision nodes from multiple availability zones on the infrastructure provider to ensure a deep pool of spot instances. Distributing across availability zones does not degrade the performance of the algorithm because there is low communication between nodes during the calculation. While this multi-cluster design allows for recruitment on a larger scale, it adds complexity to the queuing and monitoring systems, where each cluster's queue must be periodically re-balanced so that all clusters remain active.

While the scale of node recruitment was addressed by the previous modifications, the speed of recruitment also needed to be improved in order to reach large scale queuing in a short time. For this, the Karpenter open-source autoscaler (https://karpenter.sh) was used in lieu of the standard Kubernetes cluster autoscaler. Karpenter removes the need for an autoscaling group, allowing for node provisioning and removal to occur at a rapid rate. An additional benefit of changing to Karpenter was the ease of scheduling multiple workers on a variety of nodes, which improved spot instance availability across all clusters due to diversification of instance type and size.

In order to efficiently store the outputs from a massive number of worker nodes, Amazon Aurora was used in a "serverless" fashion with a highly scalable proxy in between (RDS Proxy) that allows pooling and sharing of the database connections between nodes. This allowed us to handle large numbers of simultaneous connections (e.g. when 62,500 or more workers are active). We optimized communication between nodes and the database by decreasing their size and frequency, preventing data transfer from becoming a hindrance to computation. Additionally, leveraging 1 million simultaneous vCPUs required a widened IP address range, achieved by deploying IPv6 networking for all components, which is novel for some of the components used.

All subproblems of order $n>1$ were run on spot instances and were occasionally evicted prior to completing. We therefore implemented a mechanism to automatically re-queue interrupted subproblems, where each is maintained by a decomposition worker that reschedules them in the event that a compute node stops before completing.

\subsection*{Performance on one million vCPUs}

Each PFAS studied in this work involved the computation of largely different numbers of $n$-body terms as summarized in Table~\ref{tbl:problemsize} and different levels of parallelism were therefore targeted for each molecule. For the dissociation curve of TFA, a single cluster with 1,000 simultaneous workers (16,000 vCPUs) was used. The larger number of terms in PFBA required 3 clusters and 5,000 workers (80,000 vCPUs). Finally, the enormous number of terms from PFOA demanded 13 clusters of approximately 62,500 workers (1,000,000 vCPUs), as discussed in the previous section. For both PFBA and PFOA, the energies of only the equilibrium and dissociated geometries were calculated in order to compute the BDE rather than the full dissociation curve. For brevity, we will only discuss the PFOA calculation in detail.

\begin{table}[ht]
\begin{center}
\begin{minipage}{\textwidth}
\caption{Problem size for each PFAS studied in this work (TFA, PFBA, and PFOA). Each molecule is roughly one order of magnitude higher in complexity that the previous, giving us three distinct stages for boosting the scalability of our software. The number of $n$-body terms are given for 4 orders of the expansion, which determine the maximum number of subproblems in each case. $n=3$ and $4$ utilized a screening procedure to compute significantly fewer terms than the listed values. Details are available in the Supplementary Information.}\label{tbl:problemsize}%
\begin{tabular}{lccccccc}
\toprule
\multirow{2}{*}{PFAS}  & Valence & Active & FCI & \multicolumn{4}{c}{Total iFCI Terms}\\
 & Electrons & Orbitals & Determinants & $n=1$   & $n=2$  &  $n=3$  &  $n=4$ \\ 
 \midrule
TFA  & 42  & 96  & $6.1\times10^{41}$ & 21 & 210 & 1330 & 5985 \\
PFBA & 78  & 174 & $1.4\times10^{78}$ & 39 & 741 & 9139 & 82251 \\
PFOA & 150 & 330 & $1.2\times10^{151}$ & 75 & 2775 & 67525 & 1215450 \\
\bottomrule
\end{tabular}
\end{minipage}
\end{center}
\end{table}

We performed iFCI up to the 4-body expansion to ensure high accuracy\cite{Rask_porphyrin,Zimmerman_iCAS}, with each $n$ performed sequentially, pausing for analysis in between. iFCI generates a combinatorially larger number of subproblems with increasing order of $n$ and only the 4-body calculation generated enough terms to require the full use of the deployed architecture. The 1-body, 2-body, and 3-body terms for PFOA were solved using 50, 100, and 500 workers, respectively. The combined count of 4-body terms for both equilibrium and dissociated geometries totalled to 416,857 subproblems, which were solved on over 62,500 simultaneous workers.

\begin{figure}[ht]
\centering
\includegraphics[width=0.8\textwidth]{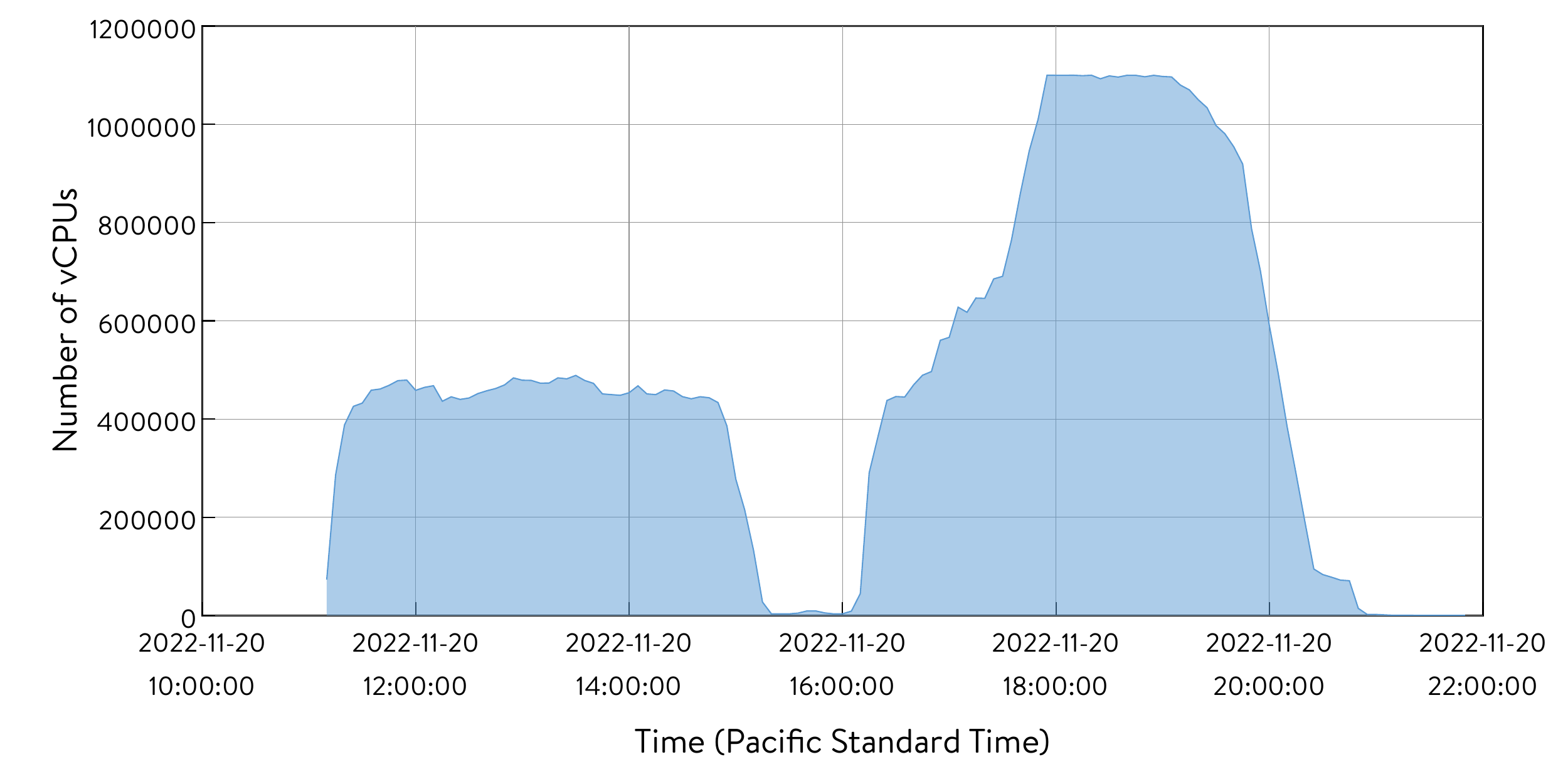}
\caption{Scalability of QEMIST Cloud during the one million vCPU run. The shaded area represents the number of recruited vCPUs across all clusters as a function of time for the majority of $n=4$ terms.} 
\label{fig:1M}
\end{figure}

The performance of QEMIST Cloud’s implementation of iFCI at this scale is shown in Figure \ref{fig:1M}. During this large scale run, more network interfaces were requested than were available,  limiting the number of nodes that were able to be recruited and  resulting in an early maximum of approximately 500,000 vCPUs. The calculation was suspended to allow for removal of many network interfaces, resulting in a gap in active number of vCPUs seen in Figure \ref{fig:1M}. Once this issue was resolved, our infrastructure was able to recruit more than one million vCPUs to simultaneously solve the 4-body iFCI subproblems. A minority of subproblems (14\%) were unable to complete due to internal errors, so those subproblems were subsequently rerun on a smaller infrastructure, not shown in Figure \ref{fig:1M}.

Timing information for all $n$-body terms for PFOA is provided in Table \ref{table:timings}. The average runtime of each subproblem increases at higher $n$ due to the higher number of orbitals being correlated. The exception to this is the 1-body fragments whose timings include a CISD calculation required to form summation natural orbitals (details in Methods section), which contributes significantly to the subproblem runtime. The wall time for completion of each fragment is summarized as well, though each geometry was run simultaneously. The full wall time for solving the 416,857 subproblems (including both geometries of PFOA) totals to 30.73 hours. The first 3.36 hours were a small test of the infrastructure ($<$0.1\% of subproblems), 9.41 hours were the large-scale run, (86\% of subproblems), and the remaining errored problems (14\%) took 17.96 hours to complete. The discrepancy in times is due only to the difference in scale of infrastructure for each of these sections; only the large-scale run is representative of the performance at one million vCPUs. These wall times exclude any periods when the infrastructure was inactive.

\begin{table}
\begin{center}
\begin{minipage}{\textwidth}
\caption{Details of the compute time for two geometries of the PFOA molecule at each order of the $n$-body expansion. Only times while the cluster was active were considered. Each geometry was run simultaneously, and therefore the wall time does not reflect the duration if they were individually run. Values for $n=1$ include the time spent running CISD to generate the one-body RDM (for each 1-body term) needed for the construction of the SNOs.}
\label{table:timings}%
\begin{tabular}{@{}llllll@{}}
\toprule
C$-$F Distance & Increment & $n=1$ & $n=2$ & $n=3$ & $n=4$ \\
\midrule
\multirow{3}{4em}{1.3\AA} & Wall time (hours) & 5.29 & 0.24 & 0.94 & 24.42\footnotemark[1] \\
 & Average runtime (seconds) & 1718.55 & 53.68 & 111.31 & 1023.11 \\
 & Total compute time (hours) & 35.80 & 41.38 & 1085.85 & 62145.66 \\
 \midrule
\multirow{3}{4em}{6.0\AA} & Wall time (hours) & 3.16 & 0.16 & 1.07 & 18.86\footnotemark[1] \\
 & Average runtime (seconds) & 1740.60 & 51.08 & 109.66 &  1257.20 \\
 & Total compute time (hours) & 36.26 & 39.37 & 982.20 & 69211.46 \\
\bottomrule
\end{tabular}
\footnotetext[1]{These simulations ran simultaneously.}
\end{minipage}
\end{center}
\end{table}

\subsection*{PFAS bond-breaking simulation}

Prior theoretical studies on PFAS primarily use density functional theory (DFT) to describe bond breaking processes. However, the BDE calculated by DFT varies significantly for different functionals\cite{Bentel_2019, Filho_2020}. High-accuracy electronic structure calculations have been limited to small molecules, as many multi-reference methods scale unfavorably with system size\cite{Filho_2020}. Additionally, the “gold standard”, CCSD(T) (coupled cluster with single and double excitations with perturbative triples), does not accurately treat stretched bonds or diradical systems\cite{NPE_1,NPE_3,Ge_2007,Lyakh_2012}. Considering studies outside of PFAS, one of the largest multireference electronic structure calculations to date is the generation of a rigid potential energy curve for breaking pentyldiazene's N=N bond\cite{He_2020}, with the number of determinants from a complete active space being roughly $10^{60}$ times smaller than PFOA’s valence space in the present study. Overall, a higher level of theory is required to treat the strong correlations involved in the bond
breaking of PFAS, and a more scalable framework is needed
to handle larger system sizes. iFCI provides this framework, being highly parallelizable and providing smooth potential energy curves for bond dissociation\cite{Zimmerman:2017ab, Rask_porphyrin}.

\begin{figure}[ht]
\centering
\includegraphics[width=1\textwidth]{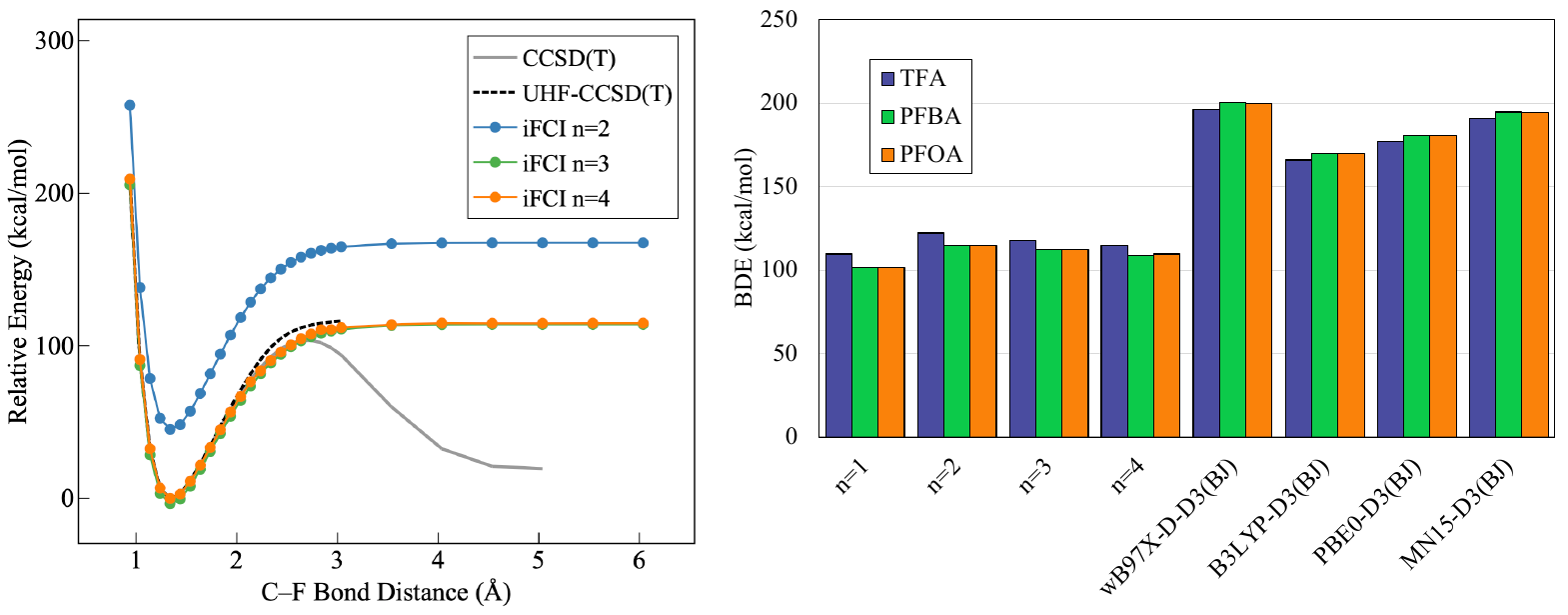}
\caption{Potential energy surface of TFA C$-$F dissociation (left) and bond dissociation energies for all three PFAS from $n=1$-4 and various DFT functionals (right).}
\label{fig:pes_bde}
\end{figure}

We computed the electronic energies at the equilibrium and stretched geometries for three PFAS with iFCI up to the 4-body expansion. This provides an accurate bond dissociation energy for the rigid-body stretch of the bond between a fluorine and the $\alpha$-carbon on the carboxylic acid group, among other chemical properties. We calculated the full potential energy surface along the bond dissociation coordinate of TFA as shown in Figure~\ref{fig:pes_bde}, and the equilibrium and dissociated geometries of PFBA and PFOA, as summarized in Table~\ref{tbl:bde}. The BDEs of each species show a similar trend with increasing $n$. The relatively small change in BDE from $n=3$ to $4$ reflects good convergence within iFCI. PFBA and PFOA’s nearly identical BDEs indicate that additional CF$_2$ units do not meaningfully affect the dissociation of a distant C$-$F bond. This does not imply that the overall PFAS length has no effect on BDE. Rather, the local geometry in these two molecules is similar for the chosen dissociated bond, and the electronic environment is therefore similar as well. Although the DFT BDE between PFBA and PFOA are similarly close, they vary by up to 30 kcal/mol between functionals. Since bond breaking is barrierless, BDE and activation energy are equivalent, and the difference between computational methods can be contextualized in terms of half-life using the Arrhenius equation\cite{Bevington_1970,Laidler_1984,Altarawneh_2022}. For example, the 90 kcal/mol difference between DFT ($\omega$B97x-D) and iFCI ($n=4$) for PFOA equates to a predicted half-life of bond breaking to be $10^{44}$ times longer for DFT compared to iFCI. The significant overestimation of the strength of the C$-$F bond by DFT may impact the study and design of PFAS remediation strategies.

\begin{table}[ht]
\begin{center}
\begin{minipage}{\textwidth}
\caption{PFAS bond dissociation energies (kcal/mol). The change in BDE with increasing $n$ of iFCI is consistent between each PFAS while the BDE varies greatly between DFT functionals.}\label{tbl:bde}%
\begin{tabular}{lcccc}
\toprule
\multicolumn{2}{l}{Method}         & TFA   & PFBA  & PFOA  \\
\midrule
\multirow{4}{*}{iFCI}    & $n=1$    & 109.6 & 101.8 & 101.7 \\
                         & $n=2$    & 122.2 & 114.5 & 114.4 \\
                         & $n=3$    & 117.5 & 112.1 & 112.4 \\
                         & $n=4$    & 114.8 & 108.7 & 109.6 \\
\multicolumn{2}{l}{$\omega$B97x-D} & 196.0 & 200.2 & 199.9 \\
\multicolumn{2}{l}{B3LYP-D3(BJ)}   & 165.9 & 170.0 & 169.7 \\
\multicolumn{2}{l}{PBE0-D3(BJ)}    & 177.0 & 180.8 & 180.4 \\
\multicolumn{2}{l}{MN15-D3(BJ)}    & 190.8 & 194.6 & 194.3 \\
\bottomrule
\end{tabular}
\end{minipage}
\end{center}
\end{table}

TFA’s potential energy curve for C$-$F dissociation maintains the correct profile of a stretched bond throughout (see Figure~\ref{fig:pes_bde}). The $n=3$ and $4$ curves agree well with CCSD(T) and UHF-CCSD(T) around the equilibrium region (e.g. $n=3$ and CCSD(T) agree to within 1 kcal/mol at the equilibrium geometry), but CCSD(T) becomes unphysical upon dissociation and therefore a meaningful comparison of the BDE cannot be made. As reported in literature\cite{NPE_1, NPE_3}, the UHF-CCSD(T) method provides marked improvement over RHF-CCSD(T). However, the difference between CCSD(T) and UHF-CCSD(T) becomes large in the intermediate region as the bond begins to dissociate. The accuracy of the UHF-CCSD(T) energy improves as the bond is stretched further toward dissociation, but it was not possible to converge the UHF-CCSD equations -- and therefore also not possible to obtain UHF-CCSD(T) energies -- for geometries beyond R$_{\mathrm{C}-\mathrm{F}}=3.0${\AA}. Additionally, we were not able to obtain UHF-CCSD(T) results at any geometry for PFOA due to the high computational demands of a system of this size. For PFBA and PFOA, we performed iFCI calculations by targeting the equilibrium and dissociated geometries to determine the BDE and did not compute the full potential energy surface. We expect these species to express similar behaviour in the dissociation region when compared to TFA (see Figure~\ref{fig:pes_bde}).  

An important metric to indicate the accuracy of our iFCI calculations is the non-parallelity error (NPE), which measures the difference (maximum vs minimum error) between potential energy curves. The NPE systematically decreases with increasing order $n$ of the $n$-body expansion as shown in Table~\ref{tbl:npe}, indicating good convergence of iFCI. The NPE between $n=3$ and $n=4$ is smaller than a few kcal/mol, suggesting the 4-body expansion approaches the FCI result. The estimated error in the present iFCI at $n=4$ is at most a few percent, considering that the bond dissociation energy is on the order of 100 kcal/mol. For TFA, the components of the NPE (maximum and minimum errors) were located near the equilibrium and dissociation geometries. As such, the NPE for PFBA and PFOA may be estimated given the error at the equilibrium and dissociated geometries. They follow a similar trend with increasing $n$ to that of TFA, indicating a similar convergence of iFCI for the larger species. 

\begin{table}[ht]
\begin{center}
\begin{minipage}{\textwidth}
\caption{Non-parallelity error (NPE) analysis of PFAS potential energy curves (kcal/mol), each made with iFCI ($n=4$) as a reference. iFCI exhibits good convergence with increasing $n$. The NPE of UHF-CCSD(T) is comparatively small, but does not converge at larger bond distances. Although the generally large value for all DFT results is as a result of being compared with $n=4$, they vary greatly between both functionals and molecules.}\label{tbl:npe}%
\begin{tabular}{lcccc}
\toprule
\multicolumn{2}{l}{Method}         & TFA   & PFBA  & PFOA  \\
\midrule
\multirow{4}{*}{iFCI}    & $n=1$    & 11.7 &  6.9 &  7.9 \\
                         & $n=2$    &  9.1 &  5.7 &  4.8 \\
                         & $n=3$    &  3.6 &  3.4 &  2.8 \\
\multicolumn{2}{l}{CCSD(T)}        &  95.1 & 71.4 & 64.2\footnotemark[1]  \\
\multicolumn{2}{l}{UHF-CCSD(T)}    &  8.8 & -- & --  \\
\multicolumn{2}{l}{$\omega$B97x-D-D3(BJ)} &  80.9 & 91.2 & 90.0 \\
\multicolumn{2}{l}{B3LYP-D3(BJ)}   &  52.2 & 61.3 & 60.0 \\
\multicolumn{2}{l}{PBE0-D3(BJ)}    &  62.0 & 72.0 & 70.7 \\
\multicolumn{2}{l}{MN15-D3(BJ)}    &  75.8 & 85.7 & 84.6 \\
\bottomrule
\end{tabular}
\footnotetext[1]{DF-CCSD(T).}
\end{minipage}
\end{center}
\end{table}

The NPEs in other methods are typically tens of kcal/mol or more\cite{NPE_1, NPE_2, NPE_3}. For example, using iFCI ($n=4$) as a reference, the NPEs in four DFT functionals range 52-81 kcal/mol, and the NPE of CCSD(T) is 95 kcal/mol (see Table~\ref{tbl:npe}). The large value for CCSD(T) is due to the failure of this method in the dissociation region. Although the NPE from the available points of UHF-CCSD(T) is much smaller, the maximum error arises from the unphysical behavior in the intermediate region. Computing fully dissociated geometries with UHF-CCSD(T) is not possible due to convergence difficulties encountered in the iterative solution of the CCSD equations. Our analysis confirms that the iFCI method implemented in QEMIST Cloud is able to provide highly accurate energies for the size of molecules and number of correlated orbitals considered in this study.

In addition to an accurate description of bond breaking, iFCI provides a means to analyze the significance of each set of molecular orbital interactions on the bond dissociation energy. The correlation energy calculated for each subproblem’s combination of $n$ orbitals represents the strength to which the electrons in these orbitals interact. Similarly, the difference in correlation energy between subproblems from two separate computations indicates the degree to which the electronic state has changed from one to the other, provided the two subproblems are formed with comparable orbitals. Prior studies have utilized this concept to gauge the impact of metal-ligand interactions on the spin gap of diradical metal complexes and a model iron-porphyrin\cite{Rask_TM,Rask_porphyrin}.

\begin{figure}[ht]
\centering
\includegraphics[width=0.8\textwidth]{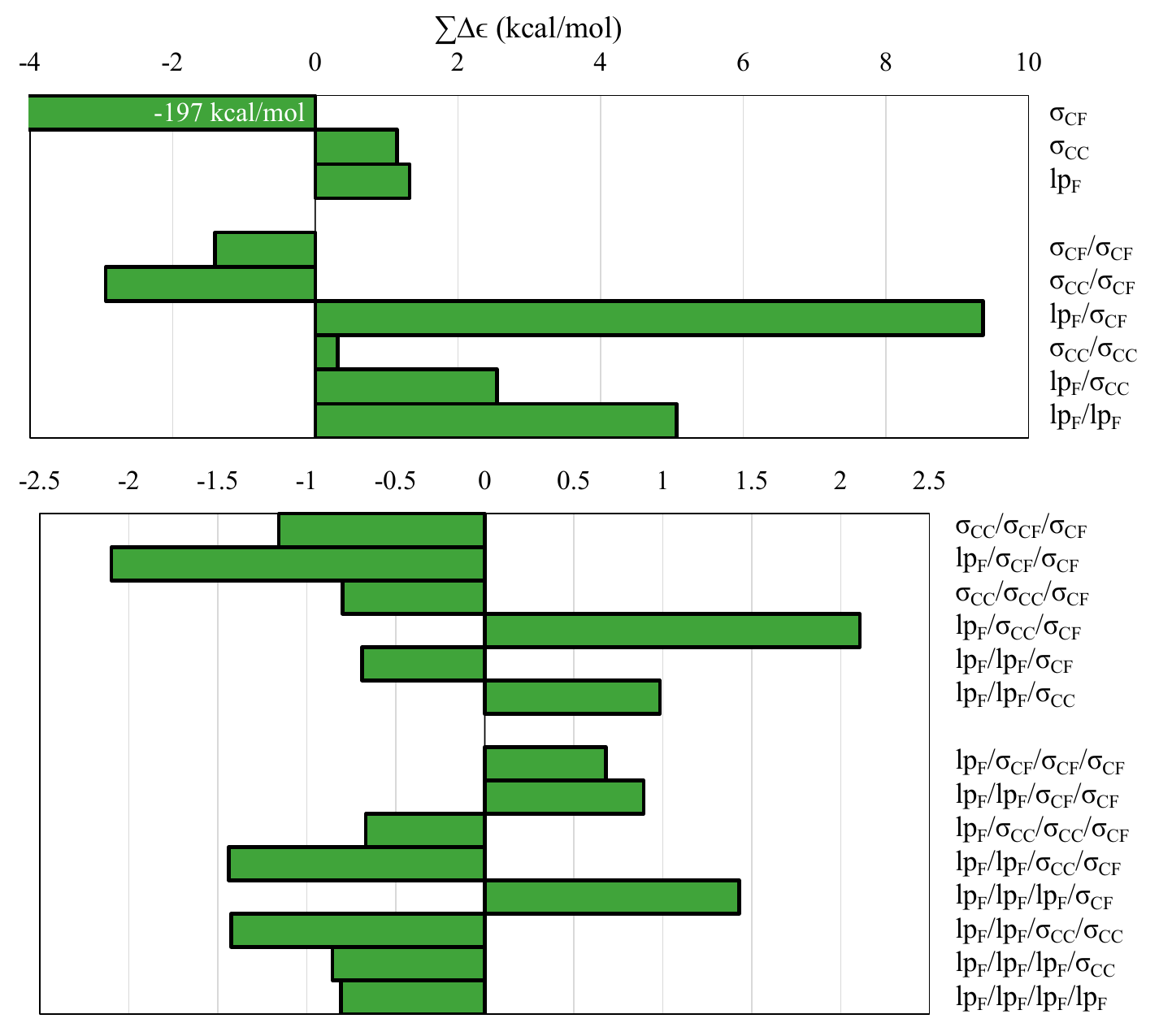}
\caption{Total contributions to the BDE from grouped combinations of orbitals for PFOA. Combinations of orbitals that result in $\lvert\sum{\Delta\epsilon}\lvert<0.5$ kcal/mol have been removed for clarity. Positive sums result in a larger overall change to the BDE (e.g. lp\textsubscript{F}/$\sigma$\textsubscript{CF} interactions cumulatively raise the BDE by $9.4$ kcal/mol).}
\label{fig:pfoa_diff}
\end{figure}

We perform a similar analysis by grouping the localized orbitals of PFOA into general characterizations with chemically relevant features: $\sigma$\textsubscript{CF}, $\sigma$\textsubscript{CC}, and lp\textsubscript{F} (orbitals present on the carboxyl group are excluded from this analysis for brevity, as they do not provide a substantial difference to the total BDE). Figure~\ref{fig:pfoa_diff} shows the contributions from the selected orbitals to the bond dissociation energy of PFOA. For example, $\sigma$\textsubscript{CF} is the contribution from 1-body terms in iFCI that correspond to the $\sigma$ bonds between C and F, and lp\textsubscript{F}/$\sigma$\textsubscript{CF} is the contribution from the 2-body terms that correspond to the interactions between one of the lone pairs of a fluorine atom and a C$-$F $\sigma$ bond. 

As expected, due to the C$-$F dissociation, $\sigma$\textsubscript{CF} has the most significant contribution ($-197$ kcal/mol) to the total bond dissociation energy, with the dissociated geometry being much lower in energy. However, in subproblems with combinations of $\sigma$\textsubscript{CF} and other orbitals, the BDE is increased by some and decreased by others. For example, the interaction between $\sigma$ bonds and the lone pairs of the fluorine atoms prefer the equilibrium structure rather than the dissociated structure (positive sum). Notable cumulative contributions to the BDE of $>1$ kcal/mol appear at $n=3$ and $4$ as well (e.g. lp\textsubscript{F}/lp\textsubscript{F}/$\sigma$\textsubscript{CC}/$\sigma$\textsubscript{CF}). This indicates that highly-excited configurations at higher $n$ significantly influence the electronic state, which are not present in other common computational methods, such as CCSD(T). Further analysis of orbital-specific contributions from the many-body expansion may provide future insights in rational design of PFAS remediation processes, such as identifying relative strengths of bonds to calculate the energetics of PFAS degradation\cite{Filho_2020,Liu_2018,Zhang_2019}.

\section*{Discussion}

We orchestrate more than one million vCPUs to execute a state-of-the-art quantum chemistry algorithm in a fully distributed modern cloud-based HPC software architecture, showcasing a massively parallel implementation of the iFCI method to tackle the bond breaking process of PFAS. We compute the rigid C$-$F bond dissociation energies (BDEs) in three different PFAS species, where the largest (PFOA) was achieved by making use of more than one million vCPUs. In addition to accurate BDEs, additional analysis of orbital contributions to these energies is made available via the separable $n$-body expansion of iFCI. These calculations constitute the first quantum chemical method to use more than one million vCPUs on the cloud and are the most accurate electronic structure calculations on PFAS species to date. Calculations of this kind provide solutions to previously intractable computational challenges and lay the foundation for further development of novel protocols for the degradation of PFAS. Potential next steps include integration of this method with other tools that are useful for the study of PFAS bond-breaking chemistry (e.g. reaction path prediction, machine learning based screening, surface reactions, consideration of environmental effects such as solvation, etc.).

The findings presented in this study illustrate that highly accurate quantum chemistry calculations are now accessible via public cloud computing infrastructure at unprecedented scalability using specifically designed algorithms and software. We believe the availability of this new HPC paradigm can dramatically increase the application of computational quantum chemistry in various fields of research and also inspire the development of new quantum chemistry algorithms more suited for highly distributed and elastic cloud computing infrastructure.

\section*{Methods}
\subsection*{Preparation of molecular geometries}
The molecular geometries of TFA, PFBA, and PFOA were optimized using the DF-MP2 method in an aug-cc-pVTZ basis set.
For the subsequent iFCI calculations, a cc-pVDZ basis set was employed.

\subsection*{Coupled-Cluster calculations}
All of the coupled-cluster calculations reported herein were performed in the Psi4 package\cite{Psi4}. The RHF-CCSD(T) calculations on PFOA were performed using the density-fitting CCSD(T) code in Psi4, which is described in Ref.~\cite{DF_CCSD_T_Psi4}. The RHF-CCSD(T) and UHF-CCSD(T) calculations for TFA and PFBA were performed without the use of density fitting. Stability analysis was employed to obtain the broken symmetry UHF reference wavefunction. 

\subsection*{DFT calculations}
All DFT calculations were performed with Psi4\cite{Psi4}, where we used the \mbox{$\omega$B97x-D}, B3LYP, PBE0 and MN15 exchange-correlation (XC) functionals in a spin-restricted formalism with the cc-pVDZ basis set and density fitting. We employed grids with 99 radial and 590 spherical points. The D3BJ dispersion correction were applied to all XC functionals except for the \mbox{$\omega$B97x-D}.

\subsection*{Incremental Full Configuration Interaction (iFCI)}
The incremental full configuration interaction approach\cite{Zimmerman:2017ab, Zimmerman:2017aa, Rask_TM, Rask_porphyrin} makes use of an incremental or many-body expansion of the correlation energy in terms of occupied orbitals labelled by $i, j, k, l, \ldots$
    \begin{align}\label{Ec_dec}
	E_\text{c}&=\sum_i \epsilon_{i} + \sum_{i>j} \epsilon_{ij} + \sum_{i>j>k} \epsilon_{ijk} +\sum_{i>j>k>l} \epsilon_{ijkl} +\ldots,
\end{align}
    in which $\epsilon_{i}$, $\epsilon_{ij}$, $\epsilon_{ijk}$,  $\epsilon_{ijkl}$ are respectively, one-, two-, three- and four-body incremental energies defined as
    \begin{align}
	\epsilon_{i} &= E_\text{c}(i)\\
	\epsilon_{ij} &= E_\text{c}(ij) - \epsilon_{i} - \epsilon_{j}\\
	\epsilon_{ijk} &= E_\text{c}(ijk) - \epsilon_{ij} - \epsilon_{ik} - \epsilon_{jk} - \epsilon_{i} - \epsilon_{j} - \epsilon_{k}\\
	\epsilon_{ijkl} &= E_\text{c}(ijkl) - \epsilon_{ijk} - \epsilon_{ijl} - \epsilon_{ikl} - \epsilon_{jkl} - \epsilon_{ij} - \epsilon_{ik} - \epsilon_{il} \nonumber \\ & - \epsilon_{jk}  - \epsilon_{jl} - \epsilon_{kl} - \epsilon_{i} - \epsilon_{j} - \epsilon_{k} - \epsilon_{l}.
\end{align}
Here, $E_c(i), E_c(ij), E_c(ijk)$ and $E_c(ijkl)$ are respectively, the correlation energies obtained from calculations including only the occupied orbital $i$, the occupied orbitals $i$ and $j$, the occupied orbitals $i, j$ and $k$ and the occupied orbitals $i, j, k$ and $l$. 

In order to make calculations tractable for larger systems, the incremental correlation energies $E_c(i), E_c(ij), E_c(ijk), \ldots$ are computed using the Heat-Bath Configuration Interaction (HBCI) approach\cite{HBCI}, which is an efficient selected configuration interaction method that constitutes as an approximation to full configuration interaction. In this method, inspired by heat bath sampling, a number of the most important determinants are selected for a variational configuration interaction treatment. The determinant space is then expanded and second-order Epstein-Nesbet perturbation theory is performed to correct for the missing correlation effects. Two parameters are used to control the selection of determinants for the variational and perturbative calculations, $\varepsilon_1$ and $\varepsilon_2$ respectively. For the calculations reported herein, we used values of $\varepsilon_1 = 1 ~\text{mHa}$ and $\varepsilon_2 = 1 ~\mu\text{Ha}$.

To further increase the efficiency of calculations without sacrificing accuracy, we make use of Summation Natural Orbitals (SNO) introduced in Ref.~\citenormal{Zimmerman:2017ab}, which facilitate a selection of the virtual orbitals for each $n$-body term and discard a number of them that have little effect on the correlation energy. Namely, for each 1-body term, the virtual space is spanned by natural orbitals of the exact CISD 1-body RDM (i.e. CISD is exact for 2-electron systems such as a 1-body term) and the natural orbitals for which the occupation numbers fall below a threshold $10^{-\eta}$ are discarded. For higher $n$, a density is formed as the sum of the one-particle CISD density matrices for each occupied orbital included in the given $n$-body term. For example, a 3-body term consisting of the occupied orbitals $i, j$ and $k$ would form a density as the sum of the 1-body densities for each 1-body term (i.e. $\rho_i, \rho_j$ and $\rho_k$, respectively)
\begin{align}
    \rho_{ijk} = \rho_i + \rho_j + \rho_k.
\end{align}
Then in analogous fashion to the 1-body case, the virtual space for the 3-body term is spanned by natural orbitals of this SNO density and the natural orbitals with occupation numbers falling below a threshold $10^{-\eta}$ are discarded. The final HBCI calculation is then performed in the truncated virtual space. Herein, we used $\eta=8.5$ as a middle ground of accuracy and efficiency, as $9.5$ did not show significantly different energies for TFA.

An additional way to reduce the computational cost of calculations without having a significant effect on accuracy is to screen certain higher order terms entering the incremental expansion of Eq.~\ref{Ec_dec}. As the size of the system increases, the number of terms entering Eq.~\ref{Ec_dec} at $n \leq 3$ starts to become large and many of the incremental energies have negligible contributions to the total correlation energy, especially as $n$ increases. Hence, we can use the connectivity arguments discussed in detail in Ref.~\citenormal{Rask_TM} and ~\citenormal{Rask_porphyrin} to discard the less important contributions to the incremental expansion for $n \geq 3$. We chose a value of $10^{-4.6}$ Ha for the screening parameter (i.e. the connected $n$-body screening procedure is applied to $n=3$ and $4$ with a value of $\mathcal{C}=10^{-4.6}$ Ha) as a good middle-ground because correlation remains to be captured at looser cutoffs, while tighter cutoffs introduce significantly more terms with negligible change to the total. Further discussions on CnB are provided in the Supplementary Information.

Our implementation of iFCI makes use of a zeroth order reference wavefunction from Generalized Valence Bond, Perfect Pairing (GVB-PP) theory\cite{GVB_1,GVB_2,GVB_3}. This constitutes one of the simplest multi-configurational wavefunctions to include static (non-dynamical) correlation and has been shown to be a good reference wavefunction for iFCI calculations on strongly correlated systems\cite{Zimmerman:2017aa} and transition metal complexes\cite{Rask_TM}. As mentioned in Ref.~\citenormal{Zimmerman:2017aa}, GVB-PP provides an excellent localized basis for the incremental expansion of Eq.~\ref{Ec_dec}, and since the PP ansatz includes correlation beyond Hartree-Fock (i.e $E_{\text{GVB-PP}} < E_{\text{HF}}$), it also provides a better starting point for the incremental expansion in iFCI. 

In our initial GVB-PP calculations on TFA, we found it difficult to converge to the same state for different geometries at larger C$-$F separation due to a discontinuity in the potential energy surface. This was mostly due to the presence of many degenerate electronic states in this region, evidenced by a change in orientation of the bonding and lone-pair orbitals on the dissociating fluorine as the bond is stretched. Therefore, we introduced a small perturbation in the form of a point charge (charge of $-1$ a.u.) along the dissociating C$-$F bond direction at a constant distance of 5 \AA \ from the F atom, which successfully corrected the orientation of the lone-pair orbital, provided a smooth potential energy surface across the full range of geometries, and changed the GVB-PP energy by less than a few mHa. The perturbation was removed in the calculation of the reference energy for the iFCI expansion -- the energy of a single determinant computed in the basis of the converged GVB-PP MOs -- and the one-electron operator used in the calculation of the single determinantal energy was unperturbed, though the converged GVB-PP MOs used for this calculation were still perturbed to retain the proper orientation of the dissociating fluorine orbitals. Similarly, the perturbation was not included in the final iFCI calculation, even though the GVB-PP orbitals used in the calculation were perturbed to retain the proper orientation of orbitals of the dissociating fluorine atom. We confirmed that this procedure had negligible effect on the final iFCI ($n=3$ and $4$) energy for the equilibrium region of the potential energy surface. This is not surprising, as for a given starting set of localized MOs (e.g. GVB, localized HF MOs, etc.), the iFCI expansion will converge to the HBCI result as $n$ is increased. This procedure was utilized for all three molecules considered in this work.

\section*{Data availability}
Atomic coordinates of the optimized geometries of TFA, PFBA, and PFOA in this work can be found in Supplementary Information.

\section*{Code availability}
The QEMIST Cloud used in this work is available as a Software as a service (SaaS) at \url{https://goodchemistry.com/qemist-cloud/}. 

\section*{Acknowledgments}
The authors are grateful to Barry Bolding at Amazon Web Service for his support for this work. The authors are also grateful to Intel for their support. The authors thank Angela Wilson at Michigan State University for the valuable discussions and insight. They thank Stephen Harper at Accenture for his technical support and Philip Ifrah at Good Chemistry for facilitating communication.

\section*{Author contributions}

C.D., H.N., A.Z., and T.Y. conceived the project. H.N., I.G. A.Z., R.P., and T.Y. coordinated the project.
A.E.R, L.H., R.W., P.K.J., and T.Y. performed quantum chemistry calculations.
A.E.R, L.H., J.T.P., and T.Y. analyzed the quantum chemistry calculation results.
L.H. improved quantum chemistry aspect of QEMIST Cloud.
S.K., D.W., N.H., A.J., and Z.M. improved the infrastructure and networking of QEMIST Cloud.
S.K., D.W., and Z.M. implemented logging and monitoring tools in QEMIST Cloud.
S.K., D.W., A.W., R.W., N.H., A.J., and R.P. optimized application code of QEMIST Cloud.
D.W. and N.H. optimized data transfer and storage in QEMIST Cloud.
S.K., D.W., N.H., A.J., and R.P. operated the QEMIST Cloud by monitoring the progress of one million core run.
M.R.H. provided support and advice regarding the QEMIST Cloud architecture.
P.M.Z provided support and advice regarding the quantum chemistry calculations.
A.E.R, L.H, D.W., A.W., R.W., R.P., and T.Y. wrote the manuscript.
All authors contributed to discussions and revision of the manuscript to its final version.

\section*{Competing interests}
The authors declare no competing interests.

\bibliographystyle{unsrtnat}
\bibliography{references} 

\includepdf[pages=-]{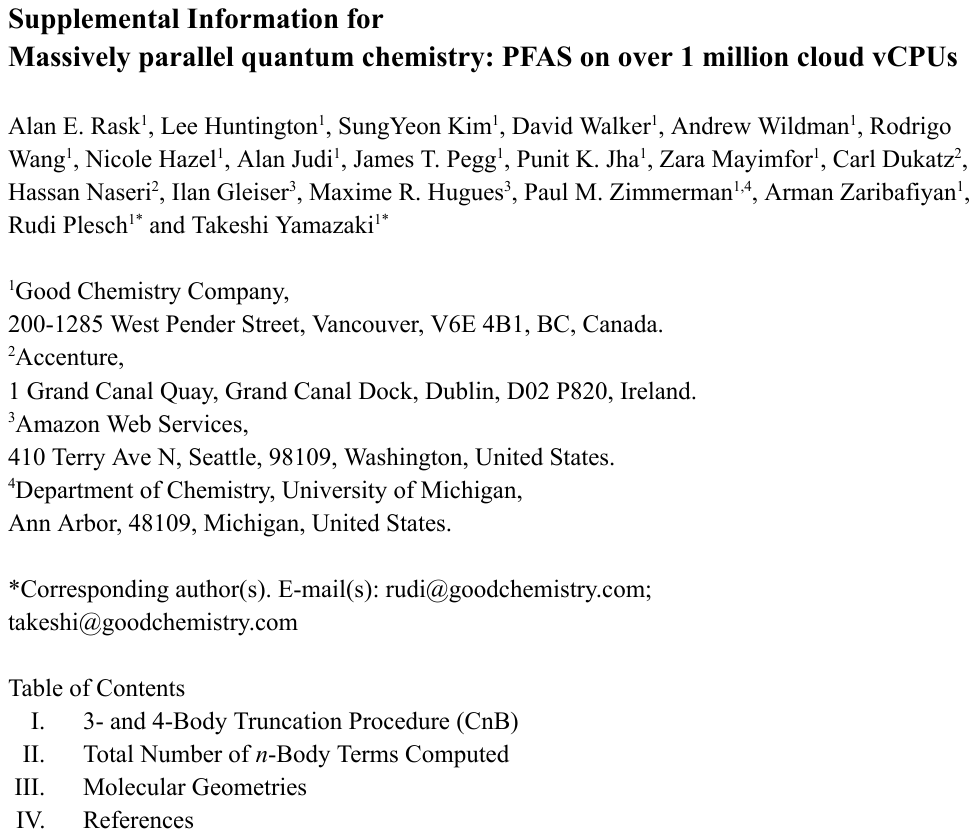}

\end{document}